# Nanocomposite RE-Ba-Cu-O bulk superconductors

**Kazumasa Iida**, Nagoya University, Department of Crystalline Materials Science, Graduate School of Engineering, Nagoya 464-8603, Japan

**Abstract** Nanocomposite oxide high-temperature bulk superconductors can be used as "quasi-magnets". Thanks to the recent progress of material processing, "quasi-magnet" with 26 mm diameter can generate a large field of 17.6 T at 26 K. These results are highly attractive for applications, involving levitation of permanent magnets on the bulk superconductors. Indeed, several other applications such as motors and magnetic resonance microscope using bulk superconductors have been proposed and demonstrated. In this chapter, we describe several techniques to improve the magnetic properties for bulk superconductors together with some basics such as phase diagrams and solidifications.

## 1. Introduction

Oxide bulk high-temperature superconductors, LRE-Ba-Cu-O [(LRE)BCO, where LRE is a light rare earth element or yttrium], are composite materials composed by a non-superconducting phase, $(LRE)_2BaCuO_5$ [(LRE)-211], and superconducting phase, $(LRE)Ba_2Cu_3O_y$ [(LRE)-123]. These materials have been commonly fabricated by the so-called top-seeded melt growth (TSMG) process which involves melting, seeding and solidifying. One of the typical applications of the bulk superconductors is a quasi-permanent magnet. When the bulk material is cooled below its superconducting transition temperature ($T_c$) in the presence of magnetic field, followed by a removal of the external fields, quantized magnetic flux can be trapped by the sample. Hence, the bulk superconductor works as a quasi-permanent magnet. The bulk superconductors can solely achieve this kind of unique applications, since their heat capacity is much higher than that of low temperature superconductors. The field trapping potential of the bulk superconductors is proportional to the product of the sample critical current density ($J_c$) and the size of the loop defined by the supercurrent. Therefore, enhancement of $J_c$ together with the fabrication of larger and weak-link free grains is a general processing aim for improving the field trapping capability of bulk superconductors. Recent progress of the technologies in sample fabrication has been remarkable and the performance of the bulk superconductors has been improved significantly. However, $J_c$ values at low field regime are typically ~50 kA/cm$^2$ at 77.3 K, which is almost two order of magnitude lower than that obtained in thin films. Optimum size of secondary phases in (LRE)BCO bulk materials, which can be obtained high $J_c$, should be similar to the twice the size of the coherence length (i.e., a few nm).

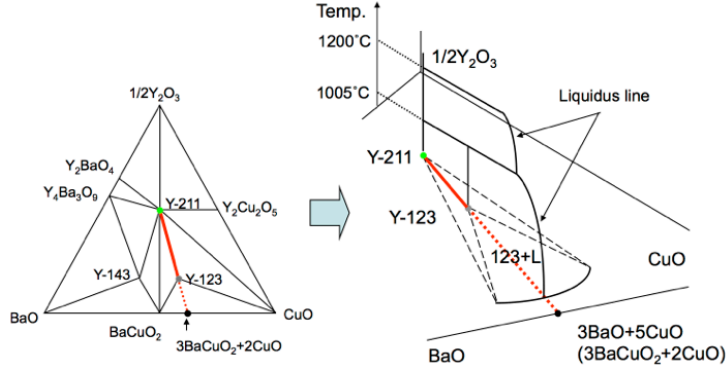

**Fig. 1.** Schematic illustration of a ternary and the corresponding pseudobinary phase diagram of Y-Ba-Cu-O system in air [1-3]

Additionally, the density of secondary phase also plays an important role for enhancing the $J_c$. However, the average size of (LRE)-211 in (LRE)-123 matrices processed by a TSMG method is in the range few tens ~ hundreds of nanometers, which leads to lower $J_c$. In this chapter, we mainly focus on how to improve the $J_c$ for (LRE)BCO bulk superconductors. Firstly, important knowledge of a (LRE)BCO bulk process involving i) phase diagram, ii) TSMG process, iii) coarsening of (LRE)-211 particles, and iv) pushing/trapping phenomena of Y-211 are reviewed in the sub-section. Then various techniques to improve $J_c$ are summarized. Finally, the importance of improving the mechanical strength of (LRE)BCO bulk samples is addressed.

## 2. Key knowledge of RE-Ba-Cu-O bulk processing

### 2-1 Phase diagram

Phase diagrams give valuable information on selecting a suitable growth method as well as its parameters. Schematic $Y_2O_3$-BaO-CuO ternary isothermal plane in air based on Ref. [1] is shown in **Fig. 1**. In solution growth, binary phase diagrams are quite useful. Hence a pseudo-binary phase diagram, which is a section through the ternary phase diagram along the tie-line between Y-211 - 3BaO+5CuO (drawn as a red line), has been intensively investigated [2-3]. There are 3 important findings in this investigation; 1) primary peritectic reaction ($Y_2O_3$ + Liquid -> Y-211) is observed at $T_{p1}$=1200 °C, 2) secondary peritectic reaction (Y-211 + Liquid ->

Y-123) is occurred at $T_{p2}$=1005 °C, and 3) yttrium solubility in Ba-Cu-O melt is quite low (i.e. 0.6 at.% at $T_{p2}$ in $pO_2$=0.21 atm). Such a low solubility of Y together with a steep liquidus line at $T_{p2}$ leads to a low growth rate of Y-123. Indeed, a typical growth rate of Y-123 phase in air has been reported around 0.05 ~ 0.3 mm/h depending on the degree of undercooling [4]. Accordingly, it takes about 20 hours to grow a YBCO bulk sample with 10 mm x 10 mm.

Unlike YBCO, (LRE)BCO systems show a relatively large LRE solubility in Ba-Cu-O liquid, a gentle slope of liquidus line at their secondary peritectic temperatures, and high peritectic temperatures [3]. As a result, the growth rate of $(LRE)_{1+x}Ba_{2-x}Cu_3O_{7-y}$ is relatively fast compared to that of Y-123. Here it is worth mentioning that $(LRE)_{1+x}Ba_{2-x}Cu_3O_{7-y}$ inherently possesses a solid solution between LRE and Ba sites and a large $x$ leads to a reduction of the $T_c$ and/or even non transition to the superconductivity [5]. In order to suppress $x$ values, oxygen controlled melt growth, OCMG, in which a whole growth process is conducted in a reduced oxygen partial pressure, has been proposed [6]. As a result, a record high $T_c$ over 96 K has been achieved in NdBCO bulk samples. Later similar high-$T_c$ has been realized in Nd-123 single crystals grown by a solute rich liquid crystal pulling, SRL-CP, method even in air process [7]. A key to reduce a solid solution level (i.e., $x$) is to employ a Ba-rich liquid. Analogous to the SRL-CP method, the employment of Ba-rich precursors in TSMG process have been also reported [8]. In all cases, finely dispersed $Nd_{1+x}Ba_{2-x}Cu_3O_{7-y}$ grains with different $x$, which work as field induced pinning centers, have been observed in the superconducting matrices [9]. As a result, field dependence of $J_c$ measurements show $J_c$ peaks at intermediate fields [10].

## 2-2 Top-seeded melt growth (TSMG) process

Appropriate amounts of $YBa_2Cu_3O_7$ (Y-123), $Y_2BaCuO_5$ (Y-211), and Pt were mixed thoroughly and then pressed into pellet. Here the addition of Pt (typically 0.1～0.5 wt.%) to the precursor powder is very effective for refinement of Y-211 in the liquid, which led to fine dispersion of Y-211 particle in the Y-123 superconducting matrix, hence high $J_c$ [11]. $CeO_2$ powder has been also used instead of Pt for cost-effective production [12]. The effect of Pt addition to the precursor pellet on microstructures will be discussed later. There have been reported two type of starting powders. One is just mentioned above, which has been commonly used to date. The other is that $Y_2O_3$ has been used instead of Y-211 [13]. In both cases, no appreciable differences have been realized in terms of final microstructures and also superconducting properties.

In order to facilitate heterogeneous nucleation and growth of a single grain, a small seed crystal has been generally used. Here it is important to mention that the seed crystals should be chemically and crystallographically compatible to the grown bulk samples. Hence Nd or SmBCO (001) melt textured or single crystals have been commonly used as a seed crystal. The seeding process is generally classified into two processes. In the first, "cold-seeding", which has been commonly used to date, involves putting the seed on the precursor pellet at room temperature. Then the pellet was heated to above the secondary $T_p$ of Y-123 phase. A fundamental limitation of the "cold-seeding" is that the secondary peritectic temperature of seed crystal should be higher than that of precursor compounds. In other words, a choice of grown materials is limited by thermal stabilities of seed crystals. However, "generic seed" with high thermally and chemically stabilities reported by Shi *et al*, (MgO-doped NdBCO melt-textured), followed by Yao *et al*, (Nd- or Sm-123 thin films on MgO substrates) opens a new avenue for growing many variety of (LRE)BCO bulk samples by the cold-seeding method [14-15].

In the "hot-seeding", green pellet is initially ramped to relatively high temperature (typically the secondary $T_p$ plus 50°C) and held at this temperature for half an hour in order to achieve a complete semi-molten state [i.e., (LRE)-211 + Liquid state]. Afterwards the semi-molten pellet is cooled to just above the secondary $T_p$. A seed crystal is then placed on the surface of the pellet prior to the crystal growth. In this method a seed crystal does not require having a higher $T_p$ than the precursor compound. However, this method needs a furnace specially designed for seeding process, in which a seed crystal is positioned on the surface of semi-molten pellet. Additionally, for OCMG process, a precise control of oxygen partial pressure during the whole process is necessary. Hence, most of the groups have been employed the "cold-seeding" to date.

Due to the high viscosity of Ba-Cu-O melt, and also co-existence of solid (LRE)-211 and liquid, the pellet is not deformed during the high temperature process. Then the sample was slowly cooled, typically at a rate of 0.3~0.5 °C/h, the crystal growth has started from the seed crystal. After the completion of crystal growth, the sample was cooled to room temperature. Due to high temperature processing, (LRE)BCO bulk is an oxygen deficient state (i.e. the sample is not superconductivity). Therefore, the sample should be annealed in $O_2$ atmosphere at around 300~400°C for 100 hours, which depends on the size of the bulk sample due to slow $O_2$ diffusion, in order to obtain the fully oxygenated bulk samples (i.e., superconducting bulk samples).

## 2-3 Coarsening of (LRE)-211 particles

As stated above in the sub-section 2-2, the addition of Pt to precursor powders is very effective for reducing the Y-211 grain growth in a partial molten state and hence fine dispersion of Y-211 in the Y-123 matrices [11]. In melt-process, average size of Y-211 particles become larger in Ba-Cu-O liquid with time according to the Ostwald ripening theory [16], which is not favorable for obtaining high $J_c$. The following equation represents the Ostwald type grain growth:

$$R^3 - R_0^3 = \alpha D_L \Gamma t / [m_L(C_S - C_0)], \tag{1}$$

where $R$ is the mean radius of Y-211 at $t$, $R_0$ is the initial mean radius, $D_L$ is the diffusivity of Y in the liquid, $\Gamma (=\sigma_{SL}/\Delta S)$ is Gibbs-Thompson coefficient, and $\sigma_{SL}$ is the interfacial energy between Y-211 and liquid. $\Delta S$ is the volumetric entropy of Y-211, $m_L$ is the liquidus slope at $T$, $C_S$ is the concentration of Y in the Y-211 phase (i.e., $C_S = 0.5$) and $C_0$ is the equilibrium liquidus composition at $R=\infty$. One can easily understand that grain growth is proportional to $t^{1/3}$. Hence the holding time in a semi-molten state should be short. Additionally, $m_L$ of Nd and Sm is larger that for Y, which leads to faster coarsening of Nd-422 and Sm-211.

It has been reported that addition of Pt to precursor powders reduces the $\sigma_{SL}$, which directly suppress the coarsening of Y-211 in liquid [16]. Almost the same effect has been reported in $CeO_2$ added YBCO [12]. It is worth mentioning that the size refinement of Y-211 is an increase in the nucleation site for Y-211. For (LRE)BCO systems, the employment of Ba-rich precursor together with $CeO_2$ addition gives also a beneficial effects on the size refinement of (LRE)-422 or Sm-211 [17-18]. Therefore Pt or Pt-related compound (or $CeO_2$) has been commonly implemented in the current melt-processing.

## 2-4 Trapping and pushing phenomena of Y-211

Macro-segregation of Y-211 particles in Y-123 matrices prepared by a TSMG method has been reported by Cima *et al*.[19], which is a similar observation pushing / trapping of foreign particles at solid-liquid interface during solidification. Later it has been reported that this segregation strongly depends on the growth rate of Y-123 phase [20]. Relatively large Y-211 particles with a small density are frequently observed in the *c*-growth sector at a small undercooling (i.e., low growth rate) [20]. In stark contrast, relatively high density of Y-211 particles has been observed in the *a*-growth sector at the same given undercooling. As stated earlier, the density of the secondary phase affects the $J_c$. Therefore, the knowledge and the understanding of such macro-segregation is needed for improving $J_c$.

On the assumption of the planar growth front of Y-123 phase as well as simplification of the following discussions, two forces may act on Y-211 particles [21]. Viscous flow around the Y-211 particles yields a drag force, $F_d$, towards the Y-123 growth front. Difference interfacial energies between solid Y-123 and Y-211 and Ba-Cu-O liquid, $\Delta\sigma_0$, as defined in equation (2):

$$\Delta\sigma_0 = \sigma_{SP} - \sigma_{LP} - \sigma_{SL} > 0, \qquad (2)$$

create a repulsive force, $F_i$, where various interfacial energies are represented by $\sigma_{SP}$ (Y-123/Y-211 particle), $\sigma_{LP}$ (Ba-Cu-O liquid/Y-211 particle) and $\sigma_{SL}$ (Y-123/Ba-Cu-O liquid) [22]. Note that a positive difference $\Delta\sigma_0$ is a necessary condition for particle pushing. A force balance between $F_d$ and $F_i$ governs macro-segregation (i.e., pushing/trapping) of Y-211 particles in Y-123 crystals.

According to the pushing/trapping theory [22], the critical Y-211 particle size for pushing, $r^*$, is given by equation (3):

$$R^* \propto \Delta\sigma_0/\eta r^*, \qquad (3)$$

where $R^*$ is the critical growth rate and $\eta$ is the melt viscosity. On the assumption of $R^* \times r^*$ =constant, anisotropic distribution of Y-211 particles may be quantitatively explained by a different growth rate as well as $\Delta\sigma_0$ between $a$- and $c$-growth sector of Y-123 phase.

Owing to the pushing effects, a total amount of secondary phase particles trapped in the superconducting matrices is much smaller than that of initial composition. Secondary phase particles are accumulated in Ba-Cu-O melt by the superconducting growth front, which leads to a non-steady state solidification of the superconducting phase. Most importantly, once the volume fraction of secondary phase particles in Ba-Cu-O liquid reaches over 50 %, the crystal growth of superconducting phase will be terminated [23].

## 3. High $J_c$ processing

### 3-1 Refinement of (LRE)-211 particles

Unlike LRE-123 type thin films, incorporation of nano-sized second phase particles such as LRE-211 in the superconducting matrices has been one of the most challenging issues in the bulk superconductors. Fortunately, there has been reported one important clue that the final size of LRE-211 particles in the LRE-123 matrices (i.e. after melt-process) has been determined by their initial size, as shown in **Fig. 2** [24]. This can be easily understood from eq. (1). Additionally, refinement of LRE-211 is an increase in the nucleation site for LRE-211, which has already been stated in the previous sub-section.

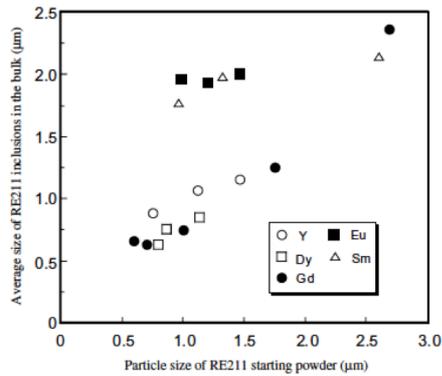

**Fig. 2.** Relationship between the particle size of RE211 starting powder and the average size of RE211 or Nd422 inclusions trapped in RE123 matrix phases (RE:Y, Dy, Gd, Eu, Sm, Nd). Reprinted with permission from Ref. 24

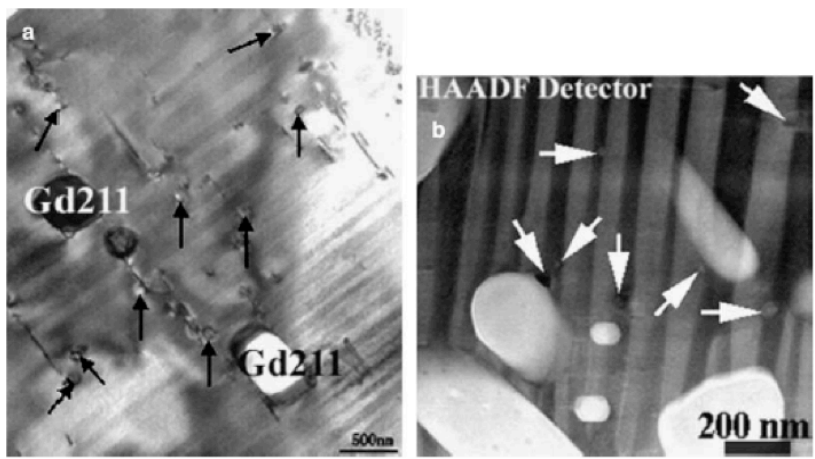

**Fig. 3.** (a) TEM micrographs of a $(Nd_{0.33}Eu_{0.33}Gd_{0.33})Ba_2Cu_3O_y$ sample with 30 mol% Gd-211 (average particle size is 70 nm); some Zr-rich nanoparticles are marked by white arrows. (b) HAADF-STEM image of a $(Sm_{0.33}Eu_{0.33}Gd_{0.33})Ba_2Cu_3Oy$ sample with 40 mol% Gd-211 (average particle size = 75 nm). Two types of nanoparticles are seen, one with size between 100 nm and 200 nm and the other with the size below 50 nm. The same contrast corresponds to the same chemical composition, white particles are Gd-rich SEG-211 and Gd-211 secondary phase. The smallest particles are $LRE-Ba_2CuZrO_y$ and $(LRE,Zr)BaCuO_y$. Reprinted with permission from Ref. 27

Hence employment of fine LRE-211 precursor powders in the melt-process has been effective in designing favorable bulk microstructures.

In order to prepare fine LRE-211 precursor powders, employing a low calcinations temperature in synthesis of LRE-211 has been proposed and indeed fine Gd-211 powders with an average size of 1.0 μm have been obtained [24]. As a result, $J_c$-$B$ characteristics of GdBCO sample prepared from such fine powder have been improved significantly.

Ball milling technique has been also a very promising way for reducing the initial size of (LRE)-211 particles [25]. This method has been quite successful in YBCO[26], GdBCO[25] and mixed (LRE)BCO[27]. For YBCO, finely dispersed Y-211 particles have been observed in Y-123 matrices [26]. As a result, a record high self-field $J_c$ of 1.1 x $10^5$ A/cm$^2$ at 77 K has been reported in melt-textured YBCO bulk samples, although inhomogeneous distribution of Y-211 particles due to the pushing effects described above. Muralidhar *et al*. have made a detail investigation on microstructures of mixed (LRE)BCO melt-textured bulk samples prepared from ball-milled (LRE)-211 precursor powders [27]. Here, LRE is an equimolar mixture of Nd, Eu and Gd. They observed nm-sized Zr-related compounds were dispersed in the (LRE)-123 matrices, as shown in **Fig. 3**. Thanks to such nano-scaled secondary phases, in-field performance has been improved significantly in mixed (LRE)BCO bulk samples.

To date, a record self-field $J_c$ of 380 kA/cm$^2$ at 77 K has been reported in (Gd,Y)BCO bulk samples with using milled Y-211 precursor powders [28]. Microstructural investigation revealed an inhomogeneous distribution of Y-211 in the superconducting matrices and hence $J_c$-$B$ performance showed a strongly position dependence, as shown in **Fig. 4**. Nevertheless, thanks to a high $J_c$, only a 25 mm sized bulk sample can trap a large field of 1.47 T at liquid Nitrogen temperature.

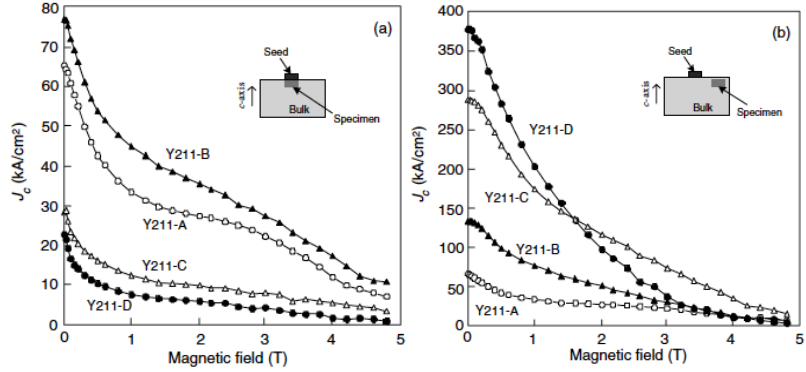

**Fig. 4.** $J_c$-$B$ curves at 77 K for the specimens cut from (a) the region near a seed crystal and (b) the upper side area of (Gd, Y)-Ba-Cu-O bulk samples fabricated with the employment of various Y211 starting powders. Reprinted with permission from Ref. 28.

## 3-2 New type of pinning centers

Many researchers have been seeking a new kind of secondary phase particles instead of Y-211 for improving $J_c$ in YBCO bulk superconductors. It has been reported that a small amount of uranium oxide leads to a formation of $YBa_2(U,Pt)O_6$[29]. Later, Hari Babu *et al*. have found a new compound, $Y_2Ba_4CuUO_y$[30]. It has been reported that U site can be replaced by Mo, W and Zr. These compounds are inert against Ba-Cu-O liquid and non-reactive with Y-123 phase. Most importantly, they do not obey Ostwald ripening theory, indicating that their size remains constant in Ba-Cu-O liquid [30]. As mentioned above, these compounds are chemically inert against Y-123. Hence the $T_c$ of YBCO containing nano-sized $Y_2Ba_4CuNbO_y$ remained a high value of over 90 K. Owing to such small secondary phase particles, $J_c$ can be improved up to 75000 A/cm$^2$ at low field at 77 K. This compound has been also used in YBCO thin films, although the compound has been changed into a simple perovskite $Ba(M_{1-x}Y_x)O_3$, where M represents Nb or Zr [31]. Nevertheless, in-field $J_c$ of such films has been improved compared to that of pristine YBCO films.

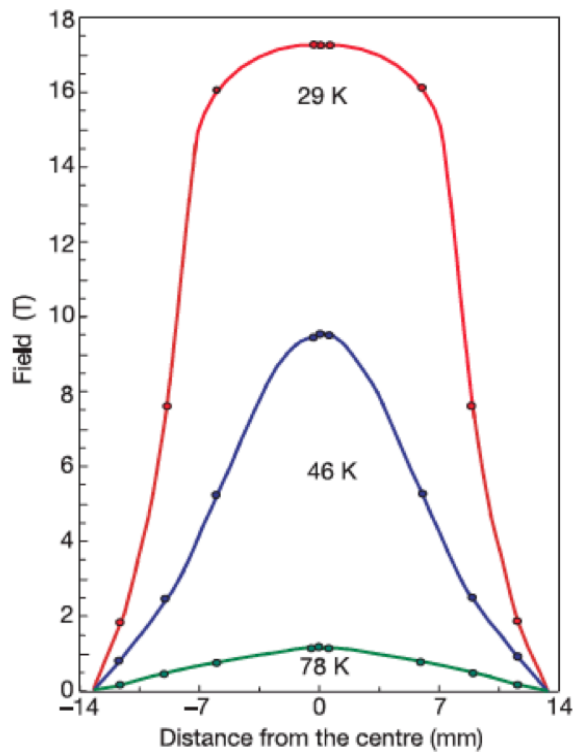

**Fig. 5.** The effect of temperature on trapped-field distribution. The field was trapped between two 26.5-mm-diameter YBCO disks with carbon fibre wrapping, resin impregnation and embedded Al. Data are shown for 29 K, 46 K and 78 K. It is evident that the trapped field is saturated at higher temperatures, but that the field is far below saturation at 29 K, showing that much higher fields could be trapped. Reprinted with permission from Ref. 35

## 4. Mechanical strength

Despite these improvements mentioned above, the level of trapped magnetic field and the ability of bulk (LRE)BCO superconductors to levitate a permanent magnet is limited severely by their mechanical strength. These materials fracture frequently when any applied electromagnetic force exceeds their relatively low tensile strength around 10~30 MPa [32]. A number of techniques have been reported to date to improve the mechanical properties of (RE)BCO bulk superconductors, including Ag addition the precursor powders, hoop stress reinforcement via the use of a metal ring and epoxy or metal impregnation under low ambient pressures. Fuchs *et al* were the first report the 2nd technique, which involved banding the periphery of a bulk YBCO sample with a stainless steel ring [33]. The reinforced sample was reported to be able to resist large applied radial tensile stresses, and a pair of YBCO disks, each of 2.6 cm in diameter, was able to trap a large field of 14.35 T at 22.5 K. The resin impregnation technique was developed initially by Tomita *et al*.[34] This involves immersing the bulk YBCO sample in a bath of liquid epoxy resin under a partial ambient vacuum and then increasing the pressure to atmospheric. Any open porosity in the sample is back-filled with liquid epoxy resin through cracks in the sample microstructure during this process. The tensile strength of epoxy back-filled YBCO increases typically by up to factor of 4 compared to the as-processed sample. Two YBCO discs impregnated with epoxy resin and containing a high thermal conductivity metal core were reported to trap a very large magnetic field of 17 T at 29 K without fracturing, as shown in **Fig. 5** [35]. Quite recently, a new record of 17.6 T at 26 K using two Ag-doped GdBCO bulks has been reported by Durrell *et al* [36].

## 5. Conclusion

Melt-processing of nano-composite (LRE)BCO bulk superconductors are reviewed briefly. Tremendous progress in enlargement of single-grain (LRE)BCO bulk samples together with high $J_c$ have been achieved. Single grain (LRE)BCO bulk superconductors with over 10 cm in diameter have already been commercially available. The ability of trapped field of GdBCO bulk superconductors exceeds 4 T at liquid nitrogen temperature. Additionally, large melt-processed single grain samples with 15 cm in diameter have been commercially available. These achievements are very attractive for applications. Indeed, the first magnetic resonance (MR) microscope using bulk superconductor magnets have been reported [37]. The inhomogeneity of the trapped magnetic field in the cylindrical region ($\phi$6.2 mm×9.1 mm) is sufficiently low value of 3.1 ppm, which allows to capture MR images of a mouse embryo.

[1] D. M. De Leeuw, C. A. H. A. Mutsaers, C. Langereis, H. C. A. Smoorenburg, P. J. Rommers, *Physica C* **152**, 39 (1988).
[2] B. J. Lee and D. N. Lee, *J. Am. Ceram. Soc.* **74**, 78 (1991).
[3] Ch. Krauns, M. Sumida, M. Tagami, Y. Yamada and Y. Shiohara, *Z. Phys. B*, **96**, 207 (1994).
[4] A. Endo, H.S. Chauhan, Y. Nakamura, K. Furuya, and Y. Shiohara, Adv. Supercond. **7**, 689 (1995).
[5] M. Murakami, N. Sakai, T. Higuchi and S. I. Yoo, *Supercond. Sci. Technol.* **9**, 1015 (1996).
[6] S. I. Yoo, N. Sakai, T. Takachi, T. Higuchi, M. Murakami, *Appl.Phys. Lett.* **65**, 633 (1994).
[7] X. Yao, M. Nakamura, M. Tagami, T. Umeda, Y. Shiohara, *Jpn. J. Appl. Phys.* **36**[2], 4A, L400 (1997).
[8] H. Kojo, S. I. Yoo, M. Murakami, *Physica C* **289**, 85 (1997).
[9] T. Egi, J. G. Wen, K. Kuroda, H. Unoki, K. Koshizuka, *Appl. Phys. Lett.* **67**, 2406 (1995).
[10] T. Higuchi, S. I. Yoo, K. Sawada, N. Sakai, M. Murakami, *Physica C* **263**, 396 (1996).
[11] N. Ogawa, I. Hirabayashi and S. Tanaka, *Physica C* **177**, 101 (1991).
[12] S. Pinol, F. Sandiumenge, B. Martinez, V. Gomis, J. Fontcuberta, X. Obradors, E. Snoeck, Ch. Roucau, *Appl. Phys. Lett.* **65**, 1448 (1994).
[13] G. Krabbes, P. Schätzle, W. Bieger, U. Wiesner, G. Stöver, M. Wu, T. Strasser, A. Köhler, D. Litzkendorf, K. Fischer, P. Görnert, *Physica C* **244**, 145 (1995).
[14] Y. Shi, N. Hari Babu, D. A. Cardwell, *Supercond. Sci. Technol.* **18**, L13 (2005).
[15] M. Oda, X. Yao, Y. Yoshida, H. Ikuta, *Supercond. Sci. Technol.* **22**, 075012 (2009).
[16] T. Izumi, Y. Nakamura and Y. Shiohara, *J. Mater. Res.* **8**, 1240 (1993).
[17] S. Matsuoka, M. Sumida, T. Umeda and Y. Shiohara, *Advances in Superconductivity X,* Springer, Berlin, 681 (1998).
[18] M. Kambara, Y. Watanabe, K. Miyake, K. Murata, Y. Shiohara and T. Umeda, *J. Mater. Res.* **12**, 2873 (1997).
[19] M. J. Cima, K. Rigby, M. C. Flemings, J. S. Haggerty, S. Honjo, H. Shen and T. M. Sung, Int. Workshop on Superconductivity, ISTEC-MRS, Maui, Hawaii, 1995, p. 55.
[20] A. Endo, H. S. Chauhan, T. Egi and Y. Shiohara, *J. Mater. Res.* **11**, 795 (1996).
[21] Y. Shiohara, A. Endo, *Materials Sci. Eng.* **R19**, 1 (1997).
[22] D. R. Uhlmann, B. Chalmers, K.A: Jackson, *J. Appl. Phys.* **35**, 2986 (1964).
[23] Y. Nakamura, M. Kambara, T. Izumi, T. Umeda, Y. Shiohara, *Sci. Technol. Adv. Mater.* **2**, 83 (2001).
[24] S. Nariki, N. Sakai, M. Murakami, I. Hirabayashi, *Physica C* **412-414**, 557 (2004).
[25] S. Nariki, N. Sakai, M. Murakami, *Physica C* **357-360**, 811 (2001).
[26] S. Nariki, N. Sakai, M. Murakami, *Supercond. Sci. Technol.* **17**, S30 (2004).
[27] M. Muralidhar, N. Sakai, M. Jirsa, M. Murakami, N. Koshizuka, I. Hirabayashi, *Physica C* **426-431**, 196 (2005).
[28] S. Nariki, N. Sakai, M. Murakami, I. Hirabayashi, *Physica C* **439**, 62 (2006).
[29] R. Sawh, Y. Ren, R. Weinstein, W. Henning, T. Nemoto, *Physica C* **305**, 159 (1998).
[30] N. Hari Babu, E. S. Reddy, D. A. Cardwell, A. M. Cambell, C. D. Tarrant, K. R. Schneider, *Appl. Phys. Lett.* **83**, 4806 (2003).
[31] E. Reich, T. Thersleff, R. Hühne, K. Iida, L. Schultz, B. Holzapfel, *Supercond. Sci. Technol.* **22**, 105004 (2009).
[32] N. Sakai, S.J. Seo, K. Inoue, T. Miyamoto and M. Murakami, 1999 *Advances in Superconductivity XI* vol. 685–688, edN Koshizuka and S Tajima (Tokyo: Springer).
[33] G. Fuchs, P. Schätzle, G. Krabbes, S. Gruß, P. Verges, K-H. Müller, J. Fink and L. Schultz, *Appl. Phys. Lett.* **64**, 2107 (2000).
[34] M. Tomita, M. Murakami and K. Katagiri, *Physica C* **378-381**, 783 (2002).
[35] M. Tomita and M. Murakami, *Nature* **421**, 517 (2003).
[36] J. Durrell, A. Dennis, J. Jaroszynski, M. Ainslie, K. Plamer, Y. Shi, A. Cambell, J. Hull, M. Strasik, E. Hellstrom, D. Cardwell, Supercond. Sci. Technol. **27**, 082001 (2014).
[37] K. Ogawa, T. Nakamura, Y. Terada, K. Kose, T. Haishi, Appl. Phys. Lett. **98**, 234101 (2011).